# The Theory of Crowd Capital

John Prpić
Beedie School of Business
Simon Fraser University
prpic@sfu.ca

Prashant Shukla
Beedie School of Business
Simon Fraser University
pshukla@sfu.ca

**Abstract**

We are seeing more and more organizations undertaking activities to engage dispersed populations through Information Systems (IS). Using the knowledge-based view of the organization, this work conceptualizes a theory of Crowd Capital to explain this phenomenon. Crowd Capital is a heterogeneous knowledge resource generated by an organization, through its use of Crowd Capability, which is defined by the structure, content, and process by which an organization engages with the dispersed knowledge of individuals –the Crowd. Our work draws upon a diverse literature and builds upon numerous examples of practitioner implementations to support our theorizing. We present a model of Crowd Capital generation in organizations and discuss the implications of Crowd Capital on organizational boundary and on IS research.

## 1. Introduction

The resource-based view (RBV) of the organization [5, 59] has demonstrated the importance of knowledge as a difficult-to-replicate resource [5, 15, 61]. From this perspective, knowledge is viewed as one of the important resources that can explain the variation in organizational capabilities and performance [37]. Building on these insights, researchers [25, 50, 51, 53, 54] have laid the foundation for what has come to be known as the knowledge-based view (KBV) of the organization.

Thus far, the KBV literature has predominantly concentrated on the mechanisms that internally develop knowledge capabilities in the organization through R&D [14] or on the internal configurations that lead to ambidextrous [38, 47], and hence adaptable, organizations. Recently however, innovation scholars have reasoned that organizations should give equal importance to internal and external knowledge for their R&D activities [11], and others have argued that the utilization of external knowledge gives organizations a competitive edge through decreased R&D costs [48]. This situation makes the KBV a pertinent and powerful lens to study Information Systems and organizations, yet, a coherent theory explaining how and why organizations engage in these disparate knowledge sources remains elusive.

In addition to this theoretical need, we note that the phenomenon of organizations undertaking activities to engage dispersed populations through the use of Information Systems (IS) continues to grow. For example, Crowdsourcing [7, 8, 29] is being widely studied in numerous contexts, and the knowledge generated from the phenomenon has been well-documented [1, 7, 30, 31, 62]. Furthermore, many organizations are using IS-tools such as "Wikis" [36] to access the knowledge of dispersed populations within the boundaries of the organization. Still other organizations are using IS-applications known as Prediction Markets, [4, 27] both internally and externally, to gather large sample forecasts.

Consequently, given this theoretical and practical impetus, this work strives to fill a vacuum in our theoretical understanding, while simultaneously situating the ever-expanding phenomena of IS applications to engage dispersed populations, such as Crowdsourcing, Wikis, and Prediction Markets, into the larger picture of IS and organizational theory.

To address these needs, we offer the theory of Crowd Capital. We conceptualize Crowd Capital as a heterogeneous organizational knowledge resource, generated by the organization's Crowd Capability: an organizational level capability that is defined by the structure, content, and process of an organizations engagement with the dispersed knowledge of individuals—the Crowd.

In this work, we present a theoretical model explaining the method of generating Crowd Capital through Crowd Capability in organizations. The implications of this unique resource are then discussed from a knowledge-based and transaction cost perspective, and from the perspective of IS theory, where in particular the theory presented here can be used to develop testable propositions regarding the centrality of IS-mediation in generating the Crowd Capital knowledge resource.

## 2. Theoretical Background

In our conception, when an organization defines the structure, content, and processes of its engagement with the dispersed knowledge of individuals, it has created a Crowd Capability, which in turn, serves to generate Crowd Capital. Figure #1 below presents a diagram for the theory of Crowd Capital that we have conceptualized. In the following subsections, we explicate each of the constructs of said theory and present examples of practitioner implementations and extant theory which support our theorizing.

**Figure #1- The Theory of Crowd Capital Creation**

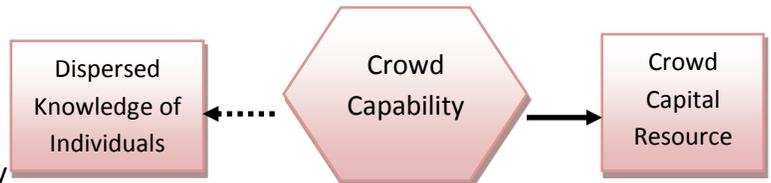

Figure #1 – The Crowd Capability of an organization engages the dispersed knowledge of individuals (through structure and content), and then generates (through internal organizational processes) a heterogeneous Crowd Capital resource.

### 2.1 Crowd Capability

We define Crowd Capability as "An organizational level capability that is defined by the structure, content, and process of an organization's engagement with the Crowd". A Crowd is any population of individuals, who supply knowledge to the organization, through Crowd Capability. A Crowd can exist inside of an organization, exist external to the organization, or a combination of the latter and the former.

The "structure" component of Crowd Capability connotes the geographical divisions and functional units within an organization, and the technological means that they employ to engage a Crowd population for the organization. As we will see in the examples given here of practitioner implementations of Crowd Capability, the structure component of Crowd Capability is always an IS-mediated phenomenon.

The "content" of Crowd Capability constitutes the knowledge, information or data goals that the organization seeks from the population. And the "processes" of Crowd Capability are defined as the internal procedures that the organization will use to organize, filter, and integrate the incoming knowledge, information, and/or data.

Furthermore, we have defined Crowd Capability as an organizational level capability, and researchers have posited that an organization's capabilities derive from its resources [37, 39]; we accept this notion, and we further contend that the particular structure, content, and processes of the Crowd Capability employed by an organization will be unique to the organization in question. This uniqueness stems from two sources; first, no two organizations have identical extant resources, and thus when creating Crowd Capability from these extant resources, the Crowd Capability created, must itself, be unique relative to other organizations. Secondly, each organization will have idiosyncratic knowledge "content" needs premised upon these same extant resources. Therefore, we would expect that Crowd Capability will to a greater or lesser degree be a unique capability for the organization.

To corroborate the conceptualization of our Crowd Capability construct, we offer numerous examples of some aspects of Crowd Capability already in use by organizations to interact with the dispersed knowledge of individuals.

Crowdsourcing, for example, is well-known as a distributed problem-solving and production model, where problems are broadcast through web-based IS to an unknown group of solvers, in the form of an open call for solutions [7]. In the case of Crowdsourcing, the structuring of heterogeneous knowledge resources is undertaken by dispersed individuals, mediated through web-based IS, "where a huge amount of individual contributions build solid and structured sources of data" [46]. In this vein, Wikipedia is perhaps the most famous example of a Crowdsourcing application.

In practitioner circles, the use of Crowdsourcing as a productive tool for firms and institutions has increased significantly in recent times [10]. Examples of Crowdsourcing deployments include these well-known examples:

- o iStockphoto[1],
- o Threadless[2],
- o Amazon's M-Turk[3],

---

[1] http://www.istockphoto.com
[2] http://www.threadless.com

- Innocentive[4],
- FBI Cryptanalysis[5]
- DARPA[6]
- GalaxyZoo[7]

As evidenced by the list above, practitioners have implemented Crowdsourcing in many diverse domains to achieve a variety of goals. The list above includes Crowdsourcing for many purposes, such as idea competitions, R&D, scientific research, problem-solving, and wage-labor pools. This variety of applications under the umbrella of Crowdsourcing in the practitioner domain would seem to indicate that the practice can be both valuable and effective as a source of knowledge for organizations. And further points to the importance of IS for Crowdsourcing, since all of the implementations listed above are mediated through web-based IS.

Furthermore, as illustrated by Majchrzak [36], we can also observe that Wikipedia (as a form of the Crowdsourcing practice) has itself spawned a new class of information systems called "Wikis". These new IS applications are being widely used by companies to instill "Crowdsourcing properties" within the organization, and highlight the value of using wikis to support organizational learning through IS.

Other IS applications such as Prediction Markets [4, 27], where organizations are using IS tools to canvass and aggregate large sample-size "predictions" from their internal employee populations, have drawn considerable attention, and have been used by diverse organizations such as the U.S. Department of Defense, Eli Lilly, General Electric, Google, France Telecom, Hewlett-Packard, IBM, Intel, Microsoft, Siemens, and Yahoo [4].

Furthermore, gamification techniques have recently emerged [16, 17] where organizations are using video-game design elements in non-game IS-contexts [18, 19] to improve user experience and user engagement with large audiences [17]. In some cases, gamification techniques are also being used to facilitate what has come to be known as Citizen Science [26, 49], including such famous IS-applications like Foldit, SETILive or GalaxyZoo.

It is our contention that the web-based IS applications found in Crowdsourcing, the IS applications known as prediction markets, the new IS applications known as Wikis, and the emerging gamification techniques using IS, are all examples of one or more dimensions of the Crowd Capability construct that we have conceptualized here. In each case, an organization creates the structure, content, and/or process to engage the knowledge of dispersed individuals through IS.

In addition to the practitioner examples illustrated above, the research community has also studied different aspects of these new Crowd Capabilities. For example, using Wikipedia as a prime example, Benkler and Nissenbaum [6] introduce the IS-mediated phenomenon of peer production, and describe it as:

> "Commons-based peer production is a socio-economic system of production that is emerging in the digitally networked environment. Facilitated by the technical

---

[3] https://www.mturk.com/mturk/welcome
[4] http://www.innocentive.com
[5] http://forms.fbi.gov/code
[6] http://challenge.gov
[7] http://www.galaxyzoo.org

infrastructure of the Internet, the hallmark of this socio-technical system is collaboration among large groups of individuals who cooperate effectively to provide information, knowledge or cultural goods without relying on either market pricing or managerial hierarchies to coordinate their common enterprise" [6]

In a somewhat similar vein, although including firms, von Hippel [57] conceptualizes innovation communities, as consisting of individuals, users, or firms interconnected by information transfer links, which may involve face-to-face, electronic, or other means of communication. Open Source Software (OSS) development is one form of an innovation community, and Linux and the Apache web server are well-known exemplars of the valuable types of products stemming from this type of community.

Similarly, Prahalad and Ramaswamy [45] introduce the idea of co-creation as "a new frame of reference for value creation" for organizations, where they state that organizations must realize that:

> "The future of competition, however, lies in an altogether new approach to value creation, based on an individual-centered co-creation of value between consumers and companies. Armed with new connective tools consumers want to interact and co-create value". [45]

It is our contention that peer production, innovation communities, and co-creation, are IS-mediated examples of theorization that illustrate some aspects of Crowd Capability. In each case, these researchers theorize some aspects of an organization creating or using a structure, content or process to engage the knowledge of dispersed individuals.

Although all of these examples illustrated here support our theorizing of Crowd Capability, our theorizing is also significantly different than any of these examples. We have parsimoniously generalized that all of these examples illustrate one or more aspects of the structure, content or process of Crowd Capability. And as we shall see in the ensuing sections, we have extended Crowd Capability as an organizational level capability that engages dispersed knowledge, therein generating the heterogeneous Crowd Capital knowledge resource.

**2.2 Dispersed Knowledge**

The existence of dispersed knowledge has been the subject of inquiry in the Economic realm for many years. Central to our understanding of dispersed knowledge is the contribution of F.A. Hayek, who in 1945 wrote a seminal work titled: "The Use of Knowledge in Society". In this work, Hayek describes dispersed knowledge as:

> "…the knowledge of the circumstances of which we must make use never exists in concentrated or integrated form but solely as the dispersed bits of incomplete and frequently contradictory knowledge which all the separate individuals possess". [28]

For Hayek, dispersed knowledge is a "body of very important but unorganized knowledge… the knowledge of the particular circumstances of time and place" [28]. Therefore, in his conception:

> "…every individual has some advantage over all others because he possesses unique information of which beneficial use might be made, but of which use can be made only if the decisions depending on it are left to him or are made with his active cooperation" [28].

For Hayek the existence of dispersed knowledge is a fact of life, not only in the realm of economics, but in all of social science:

> "The problem which we meet here is by no means peculiar to economics but arises in connection with nearly all truly social phenomena… and constitutes really the central theoretical problem of all social science. [28]

Our work accepts Hayek's view that organizations exist and compete in an environment of dispersed knowledge. Further, we contend that the through Crowd Capability an organization puts in place the structure, content, and processes to access Hayek's dispersed knowledge from individuals, each of whom has some informational advantage over the other, and thus forming a Crowd for the organization. From our perspective, it is this engagement of dispersed knowledge through Crowd Capability efforts that endows organizations with data, information, and knowledge previously unavailable to them; and the internal processing of this, in turn, results in the generation of Crowd Capital within the organization.

**2.3 Crowd Capital**

In this work, we have introduced Crowd Capital as a heterogeneous organizational knowledge resource. We label this newly emergent organizational resource as Crowd Capital because it is derived from dispersed knowledge (the Crowd), and because it is a key resource (a form of capital) for the organization that can facilitate productive and economic activity [39]. Furthermore, like the other forms of capital, Crowd Capital requires investments (for example in Crowd Capability), and potentially pays literal or figurative dividends, and hence, is endowed with typical "capital-like" qualities.

Other forms of capital have been identified in a wide variety of literatures. From a Sociological perspective, Bourdieu suggests [9] that "Capital is accumulated labor" in material or embodied form, and identifies three major categories of capital; Economic, Cultural, and Social.

More recently, Nahapiet and Ghoshal [39] investigated Social Capital and Intellectual Capital (IC) in light of Organizational Advantage. In this work they inform us that "…we use the term "intellectual capital" to refer to the knowledge and knowing capability of a social collectivity, such as an organization, intellectual community, or professional practice" and they use the term Social Capital to denote "...the sum of the actual and potential resources embedded within, available through, and derived from the network of relationships possessed by an individual or social unit."

As we can see, Intellectual Capital has been conceived as "the knowledge and knowing capability of a social collectivity, such as an organization" [39], and we contend that this construct differs significantly from Crowd Capital in two ways. First IC is a capability, and hence stems from a resource, whereas Crowd Capital is a resource that stems from a capability. Second, IC is a capability derived from a social collectivity, whereas Crowd Capital is derived from dispersed knowledge, and as we shall see below, there are significant differences between a social collectivity, and the engagement and generation of dispersed knowledge through Crowd Capability. It may be that IC is a capability that operates in parallel with Crowd Capability, and it may also be that they share some antecedent resource conditions, however, Crowd Capability defines the particular structure, content, and processes of an organization's engagement with dispersed knowledge, and hence defines a much more granular, actionable, and multi-dimensional capability. Furthermore, as we have seen with our examples of practitioner implementations of Crowd Capability, the structure of Crowd Capability is strictly an IS-mediated phenomenon, whereas IC capability can exist without IS-mediation.

On the other hand, Social Capital is generally construed to have three dimensions—structural, relational, and cognitive, [39]—which describe the "the actual and potential resources… derived from the network of relationships" [39]. This also differs significantly from our conception of Crowd Capital, because the Crowd Capability, which generates Crowd Capital, does not require a network of relationships for its engagement with dispersed knowledge, and hence Crowd Capital can be accrued without such relationship and network concerns. For example, if we consider Google's ReCaptcha system[8] for book digitization, or the famous Iowa Electronic Market[9] prediction market, these forms of IS-mediated Crowd Capability do not require a network of relationships for the accrual of Crowd Capital. This is not to say that Crowd Capability could not be leveraged to create Social Capital for an organization. It likely could, however, Crowd Capability does not require Social Capital to function.

Like several other forms of capital, some of which we have discussed here, Crowd Capital is always a heterogeneous resource. In fact, we reason that Crowd Capital is potentially a heterogeneous resource "twice-over". First of all, the Crowd Capability employed by an organization will always be idiosyncratic to the organization (either in structure, content or process, or some combination therein), based upon the idiosyncratic extant resources that form their Crowd Capability. Therefore, we would expect that since no two organizations will have identical Crowd Capability, it is also very unlikely that their Crowd Capability will thus generate a knowledge resource identical to other organizations. Second, the population of dispersed knowledge engaged by an organization's Crowd Capability is also unlikely to be the same for any two

---

[8] http://www.google.com/recaptcha
[9] http://tippie.uiowa.edu/iem/index.cfm

organizations, as illustrated by the ensuing thought experiment detailed below.

If we imagine a situation where two distinct organizations purchase the same third party IS-application to act as part of their structure of Crowd Capability (for example, some "off the shelf" web-enabled wiki platform), and we imagine that both organizations are seeking to solve the same problem (for example, "a solution to employee turnover"), first, it is highly unlikely that they manage to attract the exact same population to participate (for example, all Management Consultants in the state of New York). Second, even if such an improbable situation of attracting the exact same Crowd of participants was to arise, we would still expect that the knowledge resource that each organization generates from this effort would still be unique to the organization, because of the temporal, contextual, and emotional factors affecting the people supplying the dispersed knowledge.

Therefore, we posit that Crowd Capital will always be a heterogeneous knowledge resource for the organization. The knowledge resource garnered from Crowd Capability, may not always be "useful" initially, but it will always be unique.

In sum, Crowd Capital fits into the array of extant capitals such as social, intellectual, human, and political capital, etc., introduced previously in the social sciences and organizational research. Like many of the other forms of capital, it is a heterogeneous resource that requires investment—in this case investment in the tripartite modules of the structure, content, and processes embedded in Crowd Capability, and pays dividends as a unique organizational knowledge resource.

## 3. Discussion & Implications

We began our investigation in this work, with the observation that more and more organizations are undertaking activities to engage dispersed populations through IS. In this paper, we present theory that explains this observed phenomenon, as an attempt by organizations to engage the dispersed knowledge of individuals. Furthermore, we theorize that organizations can create a heterogeneous capability stemming from their extant resources, called Crowd Capability that will engage dispersed knowledge through particular structure, content, and processes. And that doing so, will generate a heterogeneous knowledge resource for the organization known as Crowd Capital.

While we offered various examples of how organizations are already engaging in Crowd Capability, we offer a generic and parsimonious model for Crowd Capital development within organizations and posit its workings and some potential benefits as a knowledge resource for the organization.

With the rapid advances in telecommunications in our modern business environment, such engagement with dispersed knowledge through IS, previously unreachable by the organization, becomes increasingly more efficient and tempting for organizations. Consequently, the concept of collective intelligence has been popularized as the wisdom of crowds [52], and related concepts such as crowdsourcing, prediction markets, co-creation [45], open innovation [11], and user innovation [55, 56] are being increasingly discussed among scholars.

We espouse the potential of collective intelligence – some knowledge that is more accurate when it consists of inputs from a distributed population [33]. However, we further reason that gatekeepers, the monitoring and coordination of the flow and assimilation of data, information, and knowledge, into organizations, through the Crowd Capability process, are a must for Crowd Capital to accrue. Therefore, we have strived to amalgamate the dispersed discourses on these important phenomena among scholars, and outlined a model for inculcating Crowd Capital in any organization.

While we have discussed that peer production and innovation communities represent aspects of our Crowd Capability construct, it is important to note the limits of these phenomena in relation to Crowd Capital theory. Peer production and innovation communities rely on collaboration and cooperation to exist, hence implying the need for, or the creation of, Social Capital for their operation. Further to this point, the implication of this necessity is that these efforts must be "continuing" in nature for these communities to exist and function. This is not the case for Crowd Capital. Crowd Capital can be generated through episodic and or continuing means, determined by the particular Crowd Capability created by the organization. Many of the examples that we have discussed, such as Google's ReCaptcha, the Iowa Electronic Prediction market or Foldit, illustrate the episodic nature of Crowd Capital generation.

Furthermore, examples such as Amazon's M-Turk, cannot be accounted for by the theories of peer production and innovation communities, since market pricing coordinates these efforts. Therefore, our theorization has greater explanatory power and generalizability compared to the extant theories of peer-production and innovation communities, and hence we offer an explanatory IS-theory [24] from a KBV of the organization.

Furthermore, peer production and innovation communities imply notions of autonomy in their processes, where communities self-organize and direct their collective goals and actions, and then produce outputs of some kind. These characteristics intimate that the participants in peer production and innovation communities create consensus or at the very least, have very significant input into the operation of these communities. Such input may reach as far as the actual structures by which they collaborate. In other words, participants in these communities may likely be able to alter the very structures, content, and processes by which they collaborate. With Crowd Capital, Crowd Capability efforts are guided explicitly and solely by the organization, and hence the input that the participants have into the Crowd Capability structure, content, and processes, is limited on purpose by the organization. Overall, although peer production and innovation communities share some aspects of Crowd Capability, on the whole, Crowd Capital, which is an organization-level construct and resource, has entirely different theoretical dynamics

The IS-mediated generation of Crowd Capital is also distinct from other important

IS phenomena such as data mining and business intelligence. On the one hand, Business Intelligence (BI) encompasses an array of data collection and analyses systems—online and offline—inside the enterprise, which are largely internal to the organizations and very much focussed on business processes [13, 23, 35, 58]. On the other hand, data mining, which can be thought of as a subset of BI initiatives, "…relies heavily on…techniques from machine learning, pattern recognition, and statistics to find patterns from data" [20]. While both of these IS phenomenon are important to organizations, in terms of Crowd Capital theory, they both fall squarely within the process component of Crowd Capability, potentially helping to determine the internal procedures that the organization will use to organize, filter, and integrate incoming knowledge, information, and/or data.

Crowd Capital may also have implications on organizational boundary. By invoking the dispersed knowledge of individuals through IS, an organization can expand its boundary through Crowd Capability and consequently build a unique resource base of Crowd Capital. Scholars such as Arrow and Penrose talk about the extendable nature of knowledge in an organization [3, 43], and similarly, establishing Crowd Capability in an organization, to directly seek dispersed knowledge, may also imply a blurring of the boundary of the organization with the environment. Using transaction cost theory (TCT) [12, 60] terminology, we reason that organizations are neither "making" nor "buying" the new knowledge that they seek but rather in many cases, they are relying on a crowd's benevolence in sharing their dispersed knowledge, and the organization may then employ the knowledge garnered to their local advantage. Famous examples of Citizen Science [26, 49] efforts such as Foldit[10], GalaxyZoo or SETILive[11], illustrate this blurring of organizational boundary. In some ways then, the boundary of the organization may then extend further to include all those individuals that are contributing to the organization's knowledge base. Thus, from this view, we reason that investigating the process of Crowd Capital generation from a TCT perspective, and hence explicating its implications for the boundaries of the organization, presents opportunities for research bridging the IS, KBV and TCT streams in organizational research.

Furthermore, even though we have discussed Crowd Capital at length, we have not delved into the micro-level psychological analysis of the phenomena, in terms of the individuals that are supplying the dispersed knowledge to the organization. We believe that Crowd Capital theory also provides opportunities for IS scholars to further enrich the behavioural aspects of IS theory in this respect. In particular, the question of "why a dispersed set of individuals engage with Crowd Capability through IS" remains intriguing. We forgo extensive discussion on this topic here, because the focus of our paper is on how and why organizations can augment their knowledge resources using this crowd behavior, not to explain the behavior of individuals in the crowd itself. However, we believe that this important question presents a fertile breeding ground for the cross-pollination of future IS-research with behavioural research from other realms. In particular, we reason that IS scholars have a great opportunity to inform future research, by incorporating research from public policy that uses social preference theory, [21, 22] to contradict classical economics theory, to argue that altruistic individuals behave the way that they do, because they value the utility that strangers derive from their contributions. Given the central role of IS in Crowd Capital generation it may be that this approach that we suggest is a beneficial research pathway for IS researchers.

---

[10] http://fold.it/portal/
[11] http://setilive.org/

## 4. Limitations and Future Research

Like any other research, our work is not without limitations; these can be remedied by future research. First, we believe that testing the Crowd Capital model merits further efforts from scholars. We have outlined a method for the generation of Crowd Capital and reasoned that the plethora of examples of Crowd Capability discussed in this paper provides ample opportunities for research to test the model empirically.

Second, in this work we have not delved into the ontological nature of knowledge itself. Considerable literature has done exactly this, for example [40, 41 44], where distinctions are drawn between the tacit and explicit varieties of knowledge. Similarly, others have conceptualized information and knowledge as extremes, bounding each end of a continuum [2]. Undoubtedly, this body of literature will be important in understanding the relative structure, content, processes, and overall efficacy of different Crowd Capabilities. However, our goal here is to outline the overarching theory of Crowd Capital, and we leave these interesting considerations for future research.

Third, although we have shown that elements of the structure and process components of Crowd Capability are always IS-mediated in some respect, we have not investigated the relative differences, or the specific instantiations of the structure and process of these IS-applications for Crowd Capital creation. The differing instantiations of Crowd Capability through various forms of IS are likely to have significant bearing on the amount and quality of Crowd Capital created, however, our goal here is to outline the overarching theory of Crowd Capital, and we likewise leave these very interesting considerations for future research.

And finally, following the logic of TCT, we reason that understanding the optimal level of organizational boundary expansion is important for organizations striving to generate Crowd Capital. That is, to what extent should organizations establish separate operating units for generating Crowd Capital and hence broaden their boundaries with Crowd Capability? Furthermore, what is the correct balance between investing in Intellectual Capital (such as R&D) and Crowd Capital?

Thus, while our exposition of Crowd Capital as a heterogeneous organizational knowledge resource has integrated much of the literature parsimoniously, it has also raised several intriguing questions. We hope that IS and Management scholars will see this work as an important impetus to investigate the avenues that we outline here.

## 5. Conclusion

The overarching goal of this work has been to introduce the concept of Crowd Capital as a new form of productive knowledge resource for organizations. And in doing so, explain the increasing use of IS-mediated Crowd Capability that we observe all around us in practitioner communities. To the best of our knowledge, this is the first such exposition that strives to integrate a wide array of crowd engaging phenomena into one model, which not only explicates the connection between Crowd Capability, Dispersed Knowledge, and Crowd Capital, but also situates Crowd Capital squarely into the context of the KBV.

We reason that organizations can create a unique Crowd Capability, emanating from their extant resources -comprised of some structure, content, and process- to engage with the dispersed knowledge of individuals, which is otherwise not available to them. We have also explicated how Crowd Capital is a resource arising from Crowd Capability, and is a resource that requires investment, whose benefit can be appropriated—like any other form of capital.

The theory presented here has implications for IS and Management scholars. For Management scholars using the KBV and TCT lenses in their research, this work presents opportunities to investigate individual-level and firm-level phenomena from a KBV. For IS scholars, this work presents the opportunity to investigate the panoply of IS-applications in use to create Crowd Capital, and to therefore discover the relative strengths and weaknesses of each approach. We believe that our work will help other scholars build and bound theories that do justice to the intrigue and promise that this work presents.

We hope that researchers see the value in the theory presented here, and employ it as a means to develop testable propositions, regarding not only the centrality of IS-mediation in generating Crowd Capital, but also the most efficient methods of doing so, including the consequent impacts on organizational performance.

Furthermore, we reason that Crowd Capital theory offers opportunities for cross-fertilizing IS research with other mainstream management, public policy, and economics research. While collaboration with social preference researchers might help us identify the micro-level socio-psychological explanations behind the phenomenon of crowd engagement, joining hands with management scholars might enable researchers to identify the implications of Crowd Capital on competitive advantage, organizational structure and boundary. Thus, we reason that the theory presented here potentially holds tremendous value to a wide array of researchers, particularly to IS researchers, and we hope that they see the value as we do, and follow on theses promising research avenues.

## 6. References

[1] Ågerfalk, P.J, "Outsourcing to an unknown workforce: Exploring open sourcing as a global strategy", MIS Quarterly, (32:2), 2008, pp, 385-409.

[2] Alavi, M. and D. Leidner, "Review: Knowledge management and knowledge management systems: Conceptual foundations and research issues", MIS Quarterly, (25:1), 2001, pp. 107-136.

[3] Arrow, K. J, "Economic welfare and the allocation of resources for invention." In R. R. Nelson (ed.), The Rate and Direction of Inventive Activity, pp. 609-625. Princeton, NJ: Princeton University Press, 1962

[4] Arrow, K.J, R. Forsythe, M. Gorham, R. Hahn, R. Hanson, J.O. Ledyard, S. Levmore, R. Litan, P. Milgrom, F.D. Nelson, G.R. Neumann, M. Ottaviani, T.C. Schelling, R.J. Shiller, V.L. Smith, E. Snowberg, C.R. Sunstein, P.C. Tetlock, P.E. Tetlock, H.R. Varian, J. Wolfers, and E. Zitzewitz, "The promise of prediction markets" Science, 320 (5878), 2008, pp. 877-878.

[5] Barney, J.B, "Firm resources and sustained competitive advantage", Journal of Management, (17:1), 1991, pp. 99-120.

[6] Benkler, Y. and H. Nissenbaum, (2006), "Commons-based peer production and virtue". Journal of Political Philosophy, (14), pp. 394–419.

[7] Brabham, D. C, "Crowdsourcing as a model for problem solving", Convergence, (14:1), 2008, pp. 75-90.

[8] Brabham, D, C, "Moving the crowd at threadless: Motivations for participation in a crowdsourcing application", Information, Communication & Society, (13:8), 2010, pp. 1122.

[9] Bourdieu, P, "The forms of capital", In J. Richardson (Ed.) Handbook of Theory and Research for the Sociology of Education, New York: Greenwood, 1986, pp. 241-258.

[10] Bruce, W. D, "Blogs, mashups, & wikis: Oh, my", Information Management Journal, (41:4), 2007, pp. 25-31.

[11] Chesbrough, H.W, Open Innovation: The new imperative for creating and profiting from technology, Boston: Harvard Business School Press, 2003.

[12] Coase, R.H, "The nature of the firm", Economica, (4), 1937, pp. 386-405.

[13] Cody, W. F, J.T. Kreulen, V. Krishna, and W.S. Spangler, "The integration of business intelligence and knowledge management", IBM Systems Journal (41:4), 2002, pp. 697-713.

[14] Cohen, W.M. and D.A. Levinthal, "Absorptive capacity: A new perspective on learning and innovation", Administrative Science Quarterly, (35), 1990, pp. 128-152.

[15] Conner, K.R. and C.K. Prahalad, "A resource-based theory of the firm: Knowledge versus opportunism", Organization Science; (7:5), 1996, pp. 477–501.

[16] Cooper, S, "A framework for scientific discovery through video games", Ph.D. Dissertation, University of Washington, 2011.

[17] Cooper, S, F, Khatib, A, Treuille, J, Barbero, J, Lee, M, Beenen, A, Leaver-Fay, D, Baker, Z, Popović, and Foldit players. "Predicting protein structures with

a multiplayer online game", Nature, (466:7307), 2010, pp. 756-760.

[18] Deterding, S, D, Dixon, R, Khaled, and L, Nacke, "From Game Design Elements to Gamefulness: Defining 'Gamification'", MindTrek'11, September 28-30, 2011, Tampere, Finland.

[19] Deterding, S, M, Sicart, L, Nacke, K, O'Hara, and D, Dixon "Gamification using game-design elements in non-gaming contexts", CHI 2011, May 7–12, 2011, Vancouver, BC, Canada.

[20] Fayyad , U. , G. P. Shapiro , P. Smyth , From Data Mining to Knowledge Discovery in Databases , AI Magazine , Fall 1996b , pp. 37 – 53 .

[21] Fehr, E. and U. Fischbacher, "Why social preferences matter - The impact of non-selfish motives on competition, cooperation and incentives", The Economic Journal. (112:478), 2002, pp. C1-C33

[22] Fehr, E., U. Fischbacher, and M. Kosfeld, "Neuroeconomic foundations of trust and social preferences: Initial evidence", The American Economic Review. (95:2), 2005, pp. 346-351

[23] Golfarella, M., S. Rizzi, and I. Cella, " Beyond data warehousing: what's the next business intelligence?", Proceedings of the 7[th] ACM international workshop on data warehousing and OLAP, 2004, pp. 1-6.

[24] Gregor, Shirley, "The Nature of Theory in Information Systems," MIS Quarterly, (30: 3), 2006, pp. 611-642.

[25] Gupta, A.K. and V. Govindarajan, "Knowledge Flows within Multinational Corporations", Strategic Management Journal, (21:4), 2000, pp. 473-496.

[26] Hand, E. "Citizen science: People power". Nature (466:7307), 2010, pp. 685–687.

[27] Hankins, R, and A, Lee, "Crowd sourcing and prediction markets", CHI 2011, May 7–12, 2011, Vancouver, BC, Canada.

[28] Hayek, F.A, "The use of knowledge in society", The American Economic Review, (35:4), 1945, pp. 519-530.

[29] Howe, J "The rise of crowdsourcing", Wired, (14:6), 2006, URL (accessed 05 April 2011):http://www.wired.com/wired/archive/14.06/crowds.html

[30] Horton, J. J. and L. B. Chilton, "The labor economics of paid crowdsourcing", 2010, ArXiv:1001.0627v1 [cs.HC]. 5 Jan 2010.

[31] Huberman, B.A, "Crowdsourcing and attention", Computer, (41:11), 2008, pp. 103-105.

[32] Inkpen, A. and A. Dinur, "Knowledge management processes and international joint ventures", Organization Science, (9:4), 1998, pp. 454-468.

[33] Levy, P. Collective Intelligence: Mankind's Emerging World in Cyberspace, Perseus Books: Cambridge, MA, USA, 1997.

[34] Lippman, S. A. and R. Rumelt, "Uncertain imitability: An analysis of interfirm differences in efficiency under

competition", Bell Journal of Economics, (13:2) 1997, pp. 418-438.

[35] Luhn, H. P, "A business intelligence system", IBM Journal (2:4), 1958, pp. 314-319.

[36] Majchrzak, A, "Enabling customer-centricity using wikis and the wiki way", Journal of Management Information Systems, (23:3), 2006, pp. 17-43.

[37] Makadok, R, "Toward a synthesis of the resource-based view and dynamic-capability: Views of rent creation", Strategic Management Journal; (22:5), 2001, pp. 387–401.

[38] March, J.G, "Exploration and exploitation in organizational learning", Organization Science, (2:1), 1991, pp. 71-87.

[39] Nahapiet, J. and S. Ghoshal, "Social capital, intellectual capital, and the organizational advantage", Academy of Management Review, (23:2), 1998, pp. 242-266.

[40] Nonaka, I. and G. von Krogh, "Tacit knowledge and knowledge conversion: Controversy and advancement in organizational knowledge creation theory", Organization Science (20:3), 2009, pp. 635–652.

[41] Nonaka, I, "A dynamic theory of organizational knowledge creation" Organization Science, (5:1), 1994, pp. 14-37.

[42] Nonaka, I., and H. Takeuchi, The knowledge-creating company: How Japanese companies create the dynamics of innovation, New York: Oxford University Press, 1995.

[43] Penrose, E.T, The theory of the growth in the firm. Basil Blackwell: Oxford [Eng.], 1959.

[44] Polanyi, M, The tacit dimension. London: Routledge & Kegan Paul, 1966.

[45] Prahalad, C. K. and V. Ramaswamy, "Co-Creating unique value with customers" Strategy & Leadership, (32:3), 2004, pp. 4-9.

[46] Prieur, C, D., J. Cardon, S. Beuscart, N. Pissard, and P. Pons, "The strength of weak cooperation: A case study on flickr", 2008, pp. 610-613. Retrieved from http://arxiv.org/abs/0802.2317.

[47] Raisch, S. and J. Birkinshaw, "Organizational ambidexterity: Antecedents, outcomes, and moderators", Journal of Management, (34:3), 2008, pp. 375-409.

[48] Rigby D. and C. Zook, "Open market innovation.", Harvard Business Review 80(10), 2002, pp. 80-89.

[49] Silvertown, J, "A new dawn for citizen science", Trends in Ecology & Evolution", (24:9), 2009, pp. 467-471.

[50] Spender, J.C, "Making knowledge the basis of a dynamic theory of the firm", Strategic Management Journal, (17:2), 1996, pp. 45-62.

[51] Spender, J. C. and R. M. Grant, "Knowledge and the firm: Overview", Strategic Management Journal, (17:2), 1996, pp. 5-9.

[52] Surowiecki, J, The Wisdom of Crowds. New York: Anchor Books, 2005.

[53] Szulanski, G, "Exploring internal stickiness: Impediments to the transfer of best practice within the firm" Strategic Management Journal, (17:2), 1996, pp. 27-43.

[54] Tsoukas, H, "The firm as a distributed knowledge system: A constructionist approach", Strategic Management Journal, (17:2), 1996, pp. 11-25.

[55] von Hippel, E, "The dominant role of users in the scientific instrument innovation process", Research Policy, (5:3), 1976, pp. 212-239.

[56] von Hippel, E, "Lead users: a source of novel product concepts", Management Science, (32:7), 1986, pp. 791–805.

[57] von Hippel, E., "Open source software projects as user innovation networks - no manufacturer required", In Perspectives on Free and Open Source Software, edited by J. Feller, B. Fitzgerald, S. Hissam, and K. Lakhani. Cambridge: MIT Press, 2005.

[58] Watson, H. J, and B.H. Wixom, "The Current State of Business Intelligence", Computer, (40:9), 2007, pp. 96-99.

[59] Wernerfelt, B, "A resource-based view of the firm", Strategic Management Journal, (5:2), 1984, pp. 171-180.

[60] Williamson, O.E, "Transaction cost economics: The governance of contractual relations", Journal of Law and Economics, (22:2), 1979, pp. 233-261.

[61] Winter, S, "Knowledge and competence as strategic assets", In the Competitive Challenge-Strategies for Industrial Innovation and Renewal, D. Teece (Ed.), Cambridge, MA: Ballinger, 1987.

[62] Wu, F, "Crowdsourcing, attention and productivity", Journal of Information Science, (35:6), 2009, pp. 758-765.